# Internet of Robotic Things: Current Technologies and Applications

Ghassan Samara
Computer Science Department
Zarqa University
Zarqa- Jordan
gsamarah@zu.edu.jo

Abla Hussein
Computer Science Department
Zarqa University
Zarqa- Jordan
aherzallah@zu.edu.jo

Israa Abdullah Matarneh
Computer Science Department
Zarqa University
Zarqa- Jordan
sooos_ma@yahoo.com

Mohammed Alrefai
Software Engineering Department
Zarqa University
Zarqa- Jordan
refai@zu.edu.jo

Maram Y. Al-Safarini
Computer Science Department
Zarqa University
Zarqa- Jordan
maram@zu.edu.jo

*Abstract*— **The Internet of Robotic Things (IoRT) is a new domain that aims to link the IoT environment with robotic systems and technologies. IoRT connects robotic systems, connects them to the cloud, and transfers critical information as well as knowledge exchange to conduct complicated and intricate activities that a human cannot readily perform. The pertinent notion of IoRT has been discussed in this paper, along with the issues that this area faces on a daily basis. Furthermore, technological applications have been examined in order to provide a better understanding of IoRT and its current development phenomenon. The study describes three layers of IoRT infrastructure: network and control, physical, and service and application layer. In the next section, IoRT problems have been presented, with a focus on data processing and the security and safety of IoRT technological systems. In addition to discussing the difficulties, appropriate solutions have been offered and recommended. IoRT is regarded as an essential technology with the ability to bring about a plethora of benefits in smart society upon adoption, contributing to the generation and development of smart cities and industries in the near future.**

## I. INTRODUCTION

The Internet of Things (IoT) has undergone a fast transformation and increased attention in recent years [1, 2]. IoT-based smart systems enable creating and generating a complex network of real-world devices known as 'Things' [3, 4]. Their primary function is to connect everything by exchanging and communicating data via the Internet [5], [6]. This revolutionary occurrence has had a profound impact on humanity in a variety of ways. Previously, the concept of IoT was accepted by numerous companies in every domain, such as robotics, military, nanotechnology, the healthcare sector, and space, resulting in the notion of "Internet of X Things," where X is deemed to be the suitable domain/field [7, 8, 9]. Because of the emerging concept of IoT, significant progress has been made in terms of bringing innovation in several domains such as "Internet of Cloud Things (IoCT)", "Internet of Medical Things (IoMT)", "Internet of Autonomous Things (IoAT)", "Internet of Nano Things (IoNT) [10]", "Industrial Internet of Things (IIoT)", "Internet of Mobile Things (IoMBT)", "Internet of Drone Things (IoDT).

According to Gartner [11], over 127 new IoT devices are connected to the Internet or web every second. The total number is expected to reach approximately 75 billion units by the end of 2025. It has been stated that if this entire process continued at the same rate, overall global expenditure on IoT might exceed USD 1.29 trillion this year, with a 93 percent growth in the acceptance rate of such IoT technology by organizations and corporations.

"Internet of Things (IoT)" has quickly provided a solid foundation for advancing smart industries or Industry 4.0. For their elements to function properly, smart industries rely on next-generation and sophisticatedly modern technology-based sensors and actuators [12, 13, 14, 15]. Furthermore, with the installation of the smart industry, every working personnel or employee of a company responds quickly to any alterations, modifications, or changes, which now go unnoticed.

"Industry 4.0" or "smart Industrial Internet" would serve as a link between the physical and digital worlds, which are now referred to as "Cyber-Physical Systems." Cyber-Physical Systems lay a solid foundation for the IoT phenomena, which will lead to the next cutting-edge concept in robotics known as the "Internet of Robotic Things, or IoRT" [16].

Robotics is the primary domain in which IoT technology is defining and searching for fruitful ground. This field of robotics includes a creative and fast-expanding technology that has been causing a massive and enormous transformation in many aspects of human society for more than a decade. This IoT subfield is defined as "the branch of engineering concerned with the conception, design, construction, and operation of robots" [17, 18, 19].

The uses of this sector can range from the execution of tedious and repetitive duties in the industrial and production lines to assisting and performing risky and vital operations that humans or humankind are unable to do in any way. These tasks may include, for example, executing the rescue procedure in disaster-stricken areas or completing extraterrestrial operations [20].

Previously, the machinery or robots utilized for completing or implementing these duties were essentially constituted of a single machine with numerous limitations in their hardware aspects as well as computational capacities. To address these challenges and issues, single-engine robots were first linked to a communication network by wireless or cable connections, resulting in creating a "Networked Robotic System" [21]. Despite this, they were subjected to "inherited resource limits," which resulted in network inactivity, limited memory, and minimal learning, as well as computing abilities and aptitudes [22].

At the moment, real-life scenarios necessitate speedy and intricate task implementation and execution, which requires erudite and refined data analysis as well as strong data processing abilities. The new answer to these issues has recently been handled in a modern version of "Cloud Robotics" [23, 24], which uses cloud infrastructure to gain access to resources for supporting operations. Nonetheless, these new technologies are impacted by various obstacles and issues like interoperability, security, network latency, standardization, and service quality [25, 26]. The Cloud robotics infrastructure is depicted in Fig. 1.

The "Internet of Robotic Things" excellently fits into this scenario, which strives to defy such limits by combining the domain of Robotics with the IoT, resulting in more smart, efficient, adaptive technology solutions as well as cost-effective robotic network resolution.

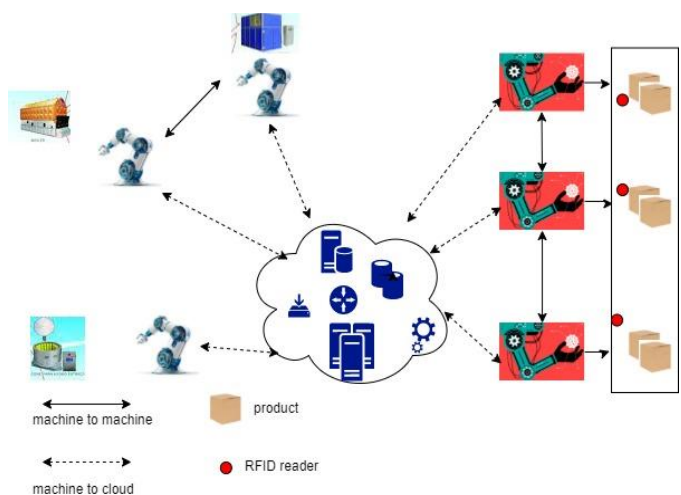

Fig. 1. Cloud Robotics infrastructure

The most notable example of Cloud Robotic-based systems that have been functional in this real-world, particularly recently, is "Self-driving or autonomous cars," such as self-driving cars launched by Google. These self-driving cars use global positioning system (GPS) navigation and are directly linked to GPS satellites as well as Google Maps; as a result, these cars receive information about their starting and ending locations, as well as real-time traffic conditions, in order to perform a precisely accurate movement or localization [27, 28].

The main advantage and benefit of IoRT technology are that it integrates two of the most important and significant modern approaches in the industry, namely, IoT and cloud robotics. Still, from another perspective, they both have to deal with many new issues and problems [29, 30, 31]. There has been relatively little research effort presented in this arena where challenging concerns have been highlighted, or solutions for combining IoT and cloud robotics or IoRT technologies have been provided. In this paper, we made an investigation by examining and analyzing more in-depth concerns and challenges that IoRT faces and comprehending them and giving answers to those challenges.

This survey aims to describe the IoRT architecture, problems, and applications that are linked with it. The primary objectives and goals of this paper are an in-depth evaluation and analysis of the "Internet of Robotic Things (IoRT)", a discussion about the IoRT infrastructure, edifying the technologies that enable IoRT and perceiving and distinguishing the currently present Robotic systems (Humanoid, Mobile, Flying, and Swarm) that have been envisioned for IoRT-based systems in the future. Another goal is to analyze and provide solutions to the significant difficulties that IoRT is now facing and solicit real-time applications that are already available for IoRT.

## II. INTERNET OF ROBOTIC THINGS – AN OVERVIEW

As previously mentioned, IoRT is thought to be the connection between IoT and the area of Robotics. This term, IoRT, was coined by Dan Kara in [32] an ABI Research report to signify and represent intelligence devices capable of keenly observing events, collecting data from various other sources as well as through sensors, and also utilizing both local and disseminated intelligence to regulate the things and objects as well as determining the appropriate eventful action to be tracked and followed.

In general, the phenomena of IoRT can be considered as a global architecture that has enabled emergent robotic services made feasible by the related connectivity of robotic things. These robotic things can benefit from the various advantages of sophisticated communication as well as interoperable tools and expertise that are based on the cloud, via IP protocol and their IPv6 version, in terms of data and information processing, computational overhead, memory storage, and, last but not least, security.

This alleviates and resolves the prior hurdles and problems encountered by the robotics domain. Furthermore, IoRT extends beyond cloud and networked-based robotic models by leveraging IoT technologies [33, 34]. This aims to integrate heterogeneous intelligent gadgets, allowing for enormous flexibility in designing and implementing distributed infrastructure that supplies computational and mathematical resources both in clouds and at the edge.

The multidisciplinary character of IoRT follows a similar trend as that of IoT. It provides enhanced and evolved robotic skills, resulting in the rise of multiple interdisciplinary solutions for different and numerous disciplines. From a technological standpoint, the IoRT enhances the traditional capabilities of a standard robot, which have been broadly categorized into three areas.

- Basic group (that involves the standpoint, manipulation, and motion)
- Higher Level group (including the autonomy, decisional, interaction cognitive)
- System-Level Group (comprising the adaptability, configurability, and dependability)

Aside from IoT features, IoRT solutions are also supported by several other techniques, such as "multi-radio access" to integrate with smart devices, Artificial intelligence (AI) to produce and create optimized and ideal solutions for intricate challenges, and cognitive technologies allowing operational proficiency.

The components of the IoRT are depicted in Fig. 2.

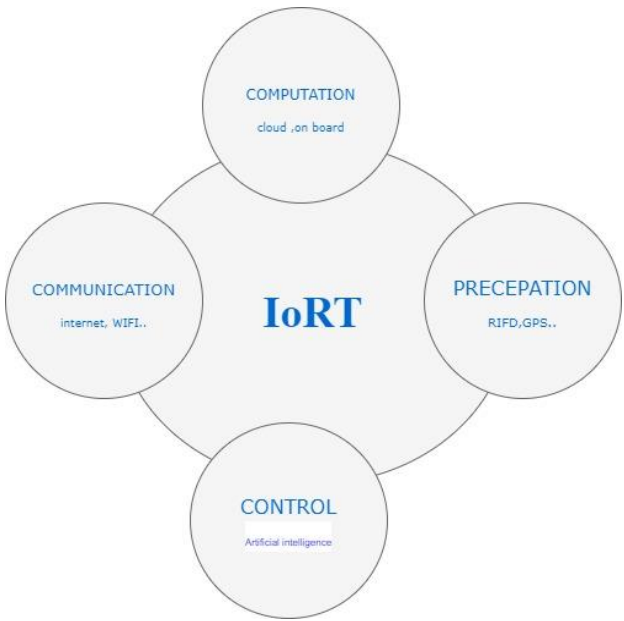

Fig. 2. IoRT components overview

To summarize, IoRT was designed and developed to be at the pinnacle of Cloud robotics, with the capabilities of combining the characteristics of IoT technology with the autonomous and self-governing nature of linked robotic objects to produce innovative and smart solutions by optimally utilizing distributed resources.

The architecture of an IoRT system is shown in Fig. 3 below. It is clear from this that its architecture is made up of three major layers [33, 34]:

- Physical Layer
- Service and Application Layer
- Network and Control Layer.

Its configuration is reminiscent of the distinctive structure of OSI architecture but with a new perspective that considers the robotic component in IoRT.

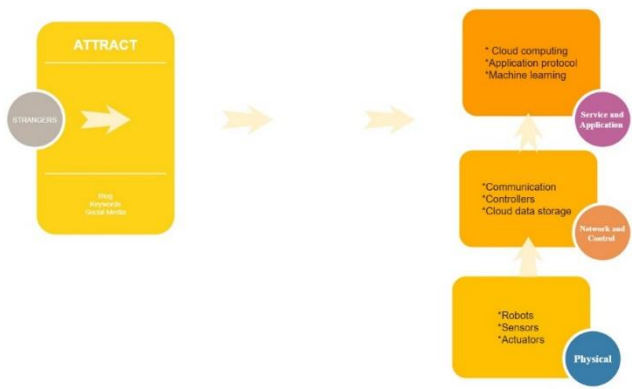

Fig. 3. Overview of IoRT architecture

## III. IORT TECHNOLOGY CHALLENGES AND SOLUTIONS

Even though IoRT technology is growing rapidly, it is still a unique domain in its early stages, with various obstacles and concerns that must be addressed. In this section, two significant problematic topics of IoRT, namely security and data processing, have been addressed, and their answers have been presented alongside.

### A. Data Processing

Because of the cyber-physical nature of IoRT technology, a great amount of data is collected in various forms from several external sources. These data gathering sources could include several types of sensors or data from the robotic mechanism or control. All of this data must be transferred from node to cloud, where it may be analyzed remotely [37]. All of this necessitates a higher level of data processing power than is typically required for IoT application processing, which leads to the following issues:

- A massive volume of data processing and streaming often necessitates the maximum communication bandwidth as well as high-tech computational capability in an IoRT network. For example, a security robot group may want to convey different audio and visual data and information at the same time when patrolling. This necessitates a swift data processing mechanism to ensure that the system responds rapidly to potential instances, events, or occurrences.

- Inactivity or latency is another difficulty. This has a significant impact on the safety and operational efficacy of real-time applications. Because numerous IoRT systems, such as "robotic transpiration systems" in smart industries, are dynamically interacting with their environment in a time-sensitive manner, an undesirable delay might cause operational failure or accidents.

- Another difficulty is unstable communication in a complex context, which can significantly limit system performance. For example, a medical system based on IoRT technology may suffer from the complicated indoor environment of a hospital building, causing interference from other medical devices such as "Magnetic Resonance Imaging (MRI)."

Possible solutions include utilizing a computational framework and optimization to address these difficulties. Computational problems can be overcome by utilizing the most recent fog and edge computing mechanisms. Both propose to distribute data storage and computing throughout the network, such that a portion of data processing can be completed near end nodes. The distinction between these is that edge computing processes a considerable quantity of data locally, whereas fog computing uses IoT gateways to process data. IoRT technology can control the use of various technologies based on the working environment and required criteria [38].

Heterogeneous IoT technology can be used to solve communication issues. It can provide a variety of selection options as well as flexibility for huge amounts of data transfer between different node types [39].

Better optimization frameworks that take pretreatment data, communication technologies, and computing nodes into account can aid in acquiring the best computational strategies, making the entire IoRT technology more proficient and successful [38].

Figure 4 depicts a comparison of the cloud computing, edge computing, and fog computing phenomena.

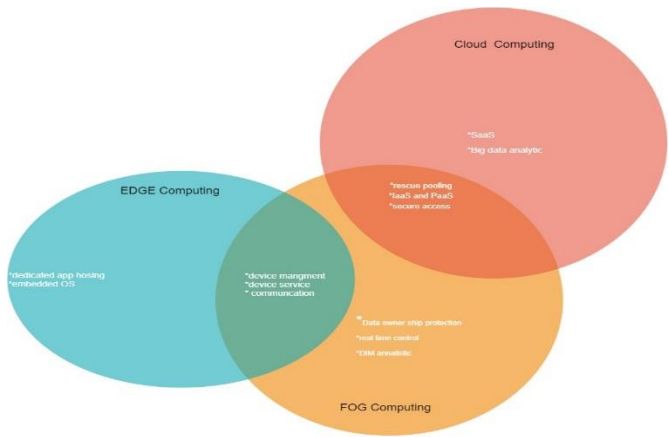

Fig. 4. Comparison between Edge, Fog & Cloud Computing

*B. Security of IoRT*

The main problem of this IoRT sector is security. Its various applications necessitate excellent safety and security, as well as stability in their functioning operations. However, the security of IoRT applications is a difficult topic in and of itself because it is linked to the security of IoT and the safety of robots. As previously said, there is a vast amount of data in IoRT technology that must be transported between the cloud and robots and between robots. This creates security and system flaws, as well as the potential for cyber-attacks by leaking critical and vital data. Some of the issues concerning IoRT security are as follows [34, 37]:

- Insecure and hazardous communication between the robot – robot or cloud – robot, resulting in the exposure of sensitive information or hacking via cyber-attack. If the connection is not sufficiently secure, adversaries or hackers discover a really simple and sophisticated approach to undermine the security of the system's safety enclosure. A specific example of this would be the hire of professional hackers by corporate competitors in order to harm other business companies.

- Another difficult issue is an authentication failure, which allows hackers or even unauthentic employees to gain simple access to the system environment by hijacking system parts, and occasionally entire systems and destroying robotic systems by reprogramming them.

- Another security risk that can arise as a result of unforeseen and sudden hardware flaws or intentional cyber-attacks is robot failure, which can sabotage the entire system environment, injure people, and create accidents.

These security concerns can be addressed and reduced by implementing the following countermeasures.

- To secure communication between two parties, encrypted communication technologies and protocols such as TLS and SSH can be utilized.

- In the future, quantum technology may be utilized to secure communication.

- In addition, numerous measures can be employed to increase authentication, such as externalized authorization, data fragments, or granular permission, which can improve authentication security.

- Furthermore, numerous new and creative technologies, such as blockchain, can be of tremendous assistance in providing communication security.

- Furthermore, robotic safety and security measures such as hardware failure analysis, "emergency detection and stop," system backup, and redundancy can be extremely beneficial in providing robustness and safety to the robotic mechanism in IoRT infrastructure.

IV. IORT TECHNOLOGY APPLICATIONS

IoRT is widely used in various sectors, and the concept of machine learning (ML) is unquestionably one of the most essential key links. To put it another way, robots associated with IoT have the ability to deliver various benefits in doing repetitive activities. They can function in catastrophe areas where humans are unable to work correctly.

*A. Disaster Response*

The researcher has deployed IoRT infrastructure for disaster response in [38]. Every year, many natural disasters such as earthquakes, typhoons, and tsunamis occur on Earth. In this particular disaster, preserving the lives of several people in the shortest amount of time is critical. In such settings or conditions, the robotic mechanism can be of considerable assistance, especially if it has been educated utilizing deep learning (DL) and artificial intelligence (AI).

The initial stage in causing this phenomenon is to deploy mobile robots to acquire information and data from the surrounding environment. A remote network expands on this data to create an AI model, which is then assessed by an internal system. This architecture is deployed from the cloud to a local workstation for performance testing. The system model is set up and positioned into robots for the next step of the learning method as soon as it reaches the appropriate dependability level. The suggested infrastructure for the disaster response application use case is depicted in Fig. 5.

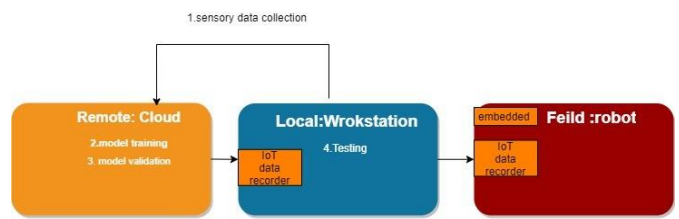

Fig. 5. IoRT Disaster Response application use case

*B. Agricultural Accuracy*

researchers in [39] produced an IoRT application for precise farming accuracy. The model design of an IoRT-based agricultural system is depicted in Fig. 6. Initially, the mobile robots collect data from their surroundings using the sensors installed onboard. This data contains information about the pressure, temperature, light, humidity, and so on. The server reclaims this information from mobile robots by processing it over Wi-Fi or cellular connectivity. The data is then installed into the online application, allowing the end-user to visualize it via the web interface.

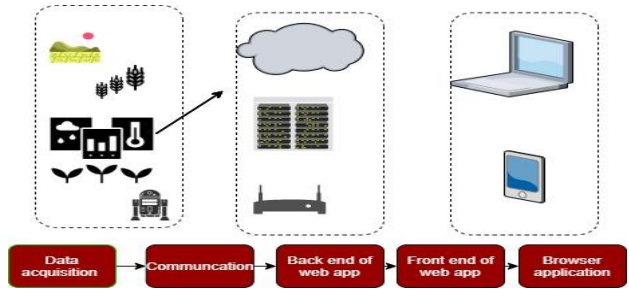

Fig. 6. IoRT Conceptual model design based agriculture mechanism

## V. CONCLUSION

The IoRT is seen as a new domain that aims to combine the IoT environment with robotic systems and technologies. IoRT connects robotic systems, connects them to the cloud, and transfers critical information as well as knowledge exchange to conduct complicated and intricate activities that a human cannot readily perform. The pertinent notion of IoRT has been discussed in this study, along with the issues that this area faces on a daily basis. Furthermore, technological applications have been examined in order to provide a better understanding of IoRT and its current development phenomenon. The study describes three layers of IoRT infrastructure: network and control, physical, and service and application layer. In the next section, IoRT problems have been presented, with a focus on data processing and the security and safety of IoRT technological systems. In addition to discussing the difficulties, appropriate solutions have been offered and recommended. IoRT is regarded as an important technology with the ability to bring about a plethora of benefits in smart society upon adoption, contributing to the generation and development of smart cities and industries in the near future.

REFERENCES

[1] Simoens, P., Dragone, M. and Saffiotti, A., 2018. The Internet of Robotic Things: A review of concept, added value and applications. *International Journal of Advanced Robotic Systems*, 15(1), p.1729881418759424.

[2] Batth, R.S., Nayyar, A. and Nagpal, A., 2018, August. Internet of robotic things: driving intelligent robotics of future-concept, architecture, applications and technologies. In *2018 4th International Conference on Computing Sciences (ICCS)* (pp. 151-160). IEEE.

[3] Samara, G. and Aljaidi, M., 2019. Efficient energy, cost reduction, and QoS based routing protocol for wireless sensor networks. *International Journal of Electrical & Computer Engineering (2088-8708)*, 9(1).

[4] Samara, G., Alsalihy, W.A.A. and Ramadass, S., 2011. Increasing Network Visibility Using Coded Repetition Beacon Piggybacking. *World Applied Sciences Journal*, 13(1), pp.100-108.

[5] "The Internet of Things: A Review of Enabled Technologies and Future Challenges." https://ieeexplore.ieee.org/document/8584051 (accessed May 18, 2021).

[6] Kumar, S., Tiwari, P. and Zymbler, M., 2019. Internet of Things is a revolutionary approach for future technology enhancement: a review. *Journal of Big data*, 6(1), pp.1-21.

[7] Samara, G. and Blaou, K.M., 2017, May. Wireless sensor networks hierarchical protocols. In *2017 8th International Conference on Information Technology (ICIT)* (pp. 998-1001). IEEE.

[8] Samara, G., Al-Okour, M. 2020. Optimal number of cluster heads in wireless sensors networks based on LEACH, *International Journal of Advanced Trends in Computer Science and Engineering, , 9(1), pp. 891–895*

[9] Samara, G. and Aljaidi, M., 2018. Aware-routing protocol using best first search algorithm in wireless sensor. *Int. Arab J. Inf. Technol.*, 15(3A), pp.592-598.

[10] Nayyar, A., Puri, V. and Le, D.N., 2017. Internet of nano things (IoNT): Next evolutionary step in nanotechnology. *Nanoscience and Nanotechnology*, 7(1), pp.4-8.

[11] "Gartner: Fueling the Future of Business," *Gartner*. https://www.gartner.com/en (accessed May 18, 2021).

[12] Samara, G., Albesani, G., Alauthman, M., Al Khaldy, M., Energy-efficiency routing algorithms in wireless sensor networks: A survey, *International Journal of Scientific and Technology Research, 2020, 9(1), pp. 4415–4418.*

[13] Samara, G., 2020, November. Wireless Sensor Network MAC Energy-efficiency Protocols: A Survey. In *2020 21st International Arab Conference on Information Technology (ACIT)* (pp. 1-5). IEEE.

[14] Alhmiedat, T. and Samara, G., 2017. A Low Cost ZigBee Sensor Network Architecture for Indoor Air Quality Monitoring. *International Journal of Computer Science and Information Security (IJCSIS)*, 15(1).

[15] Samara, Ghassan. 2020. Lane prediction optimization in VANET. *Egyptian Informatics Journal*.

[16] Damodaran, S.K. and Rowe, P.D., 2019, April. Limitations on observability of effects in cyber-physical systems. In *Proceedings of the 6th Annual Symposium on Hot Topics in the Science of Security* (pp. 1-10).

[17] Andrea, I., Chrysostomou, C. and Hadjichristofi, G., 2015, July. Internet of Things: Security vulnerabilities and challenges. In *2015 IEEE symposium on computers and communication (ISCC)* (pp. 180-187). IEEE.

[18] Alhmiedat, T.A., Abutaleb, A. and Samara, G., 2013. A prototype navigation system for guiding blind people indoors using NXT Mindstorms. *International Journal of Online and Biomedical Engineering (iJOE)*, 9(5), pp.52-58.

[19] Salem, A.O.A., Samara, G. and Alhmiedat, T., 2014. Performance Analysis of Dynamic Source Routing Protocol. *Journal of Emerging Trends in Computing and Information Sciences*, 5(2).

[20] Samara, G., Ramadas, S., Al-Salihy, W.A.H. 2010. Safety message power transmission control for vehicular Ad hoc Networks. *Journal of Computer Science,, 6(10), pp. 1056–1061.*

[21] Mohamed, N., Al-Jaroodi, J. and Jawhar, I., 2009. A review of middleware for networked robots. *International Journal of Computer Science and Network Security*, 9(5), pp.139-148.

[22] Samara, G., Al-Salihy, W.A. and Sures, R., 2010, May. Efficient certificate management in VANET. In *2010 2nd International Conference on Future Computer and Communication* (Vol. 3, pp. V3-750). IEEE.

[23] Saha, O. and Dasgupta, P., 2018. A comprehensive survey of recent trends in cloud robotics architectures and applications. *Robotics*, 7(3), p.47.

[24] Samara, G., Abu Salem, A.O. and Alhmiedat, T., 2013. Dynamic Safety Message Power Control in VANET Using PSO. *World of Computer Science & Information Technology Journal*, 3(10).

[25] Kehoe, B., Patil, S., Abbeel, P. and Goldberg, K., 2015. A survey of research on cloud robotics and automation. *IEEE Transactions on automation science and engineering*, 12(2), pp.398-409.

[26] M Khatari, G Samara, 2015, Congestion control approach based on effective random early detection and fuzzy logic, *MAGNT Research Report, Vol.3 (8). PP: 180-193.*

[27] Samara, G., Ramadas, S. and Al-Salihy, W.A.. 2010. Design of Simple and Efficient Revocation List Distribution in Urban areas for VANET's,     (IJCSIS) International Journal of Computer Science and Information Security, Vol. 8, No. 1.

[28] Samara, G., Alsalihy, W.A.H.A. and Ramadass, S., 2011. Increase emergency message reception in vanet. *Journal of applied sciences*, 11(14), pp.2606-2612.

[29] Samara, G. and Alsalihy, W.A.A., 2012, June. A new security mechanism for vehicular communication networks. In *Proceedings Title: 2012 International Conference on Cyber Security, Cyber Warfare and Digital Forensic (CyberSec)* (pp. 18-22). IEEE.

[30] Samara, G., 2018. An intelligent routing protocol in VANET. *International Journal of Ad Hoc and Ubiquitous Computing*, 29(1-2), pp.77-84.

[31] Samara, G. and Alsalihy, W.A.A., 2012. Message broadcasting protocols in VANET. *Information Technology Journal*, *11*(9), p.1235.

[32] "The Internet of Robotic Things." https://www.abiresearch.com/market-research/product/1019712-the-internet-of-robotic-things/ (accessed May 18, 2021).

[33] Samara, G., 2020. Intelligent reputation system for safety messages in VANET. *IAES International Journal of Artificial Intelligence*, *9*(3), p.439.

[34] Salem, A.O.A., Alhmiedat, T. and Samara, G., 2013. Cache Discovery Policies of MANET. *World of Computer Science & Information Technology Journal*, *3*(8).

[35] Romeo, L., Petitti, A., Marani, R. and Milella, A., 2020. Internet of robotic things in smart domains: Applications and challenges. *Sensors*, *20*(12), p.3355.

[36] Afanasyev, Ilya, Manuel Mazzara, Subham Chakraborty, Nikita Zhuchkov, Aizhan Maksatbek, Aydin Yesildirek, Mohamad Kassab, and Salvatore Distefano. "Towards the internet of robotic things: Analysis, architecture, components and challenges." In *2019 12th International Conference on Developments in eSystems Engineering (DeSE)*, pp. 3-8. IEEE, 2019.

[37] Villa, D., Song, X., Heim, M. and Li, L., 2021. Internet of Robotic Things: Current Technologies, Applications, Challenges and Future Directions. *arXiv e-prints*, pp.arXiv-2101.

[38] P. Ray, "Internet of Robotic Things: Concept, Technologies, and Challenges," *undefined*, 2016, Accessed: May 18, 2021. [Online]. Available: /paper/Internet-of-Robotic-Things%3A-Concept%2C-Technologies%2C-Ray/845c1a9a43924a5ac907861f3b98f8f32f009832

[39] Din, I.U., Guizani, M., Hassan, S., Kim, B.S., Khan, M.K., Atiquzzaman, M. and Ahmed, S.H., 2018. The Internet of Things: A review of enabled technologies and future challenges. *Ieee Access*, *7*, pp.7606-7640.

[40] Petkovic, S., Petkovic, D. and Petkovic, A., 2017. IoT devices VS. drones for data collection in agriculture. *DAAAM International Scientific Book*, *16*, pp.63-80.

[41] Dragoni, N., Giaretta, A. and Mazzara, M., 2016, May. The Internet of hackable things. In *International Conference in Software Engineering for Defence Applications* (pp. 129-140). Springer, Cham.

[42] Ray, P.P., Mukherjee, M. and Shu, L., 2017. Internet of things for disaster management: State-of-the-art and prospects. *IEEE access*, *5*, pp.18818-18835.

[43] Durmuș, H. and Güneș, E.O., 2019, July. Integration of the mobile robot and Internet of things to collect data from the agricultural fields. In *2019 8th International Conference on Agro-Geoinformatics (Agro-Geoinformatics)* (pp. 1-5). IEEE.